\documentclass[
reprint,
amsmath,
amssymb,
aip,
jcp,
floatfix
]{revtex4-2}

\usepackage{bm}
\usepackage{booktabs}
\usepackage{graphicx}
\usepackage{microtype}
\usepackage{placeins}
\usepackage{hyperref}
\hypersetup{hidelinks}

\newcommand{\avg}[1]{\left\langle #1\right\rangle}
\newcommand{\abs}[1]{\left\lvert #1\right\rvert}
\newcommand{\rr}{\mathbf r}
\newcommand{\uu}{\mathbf u}
\newcommand{\kk}{\mathbf k}

\newcommand{\Oh}{O_{\mathrm h}}

\usepackage{cancel} 
\usepackage[normalem]{ulem}
\usepackage{color}

\begin{document}

\title{Cubic-Equivariant Neural Density Functional Theory for Three-Dimensional Lattice Fluids}

\begin{abstract}
We construct a neural classical density functional that acts directly on unrestricted three-dimensional density profiles.  As a computationally tractable test bed, we consider parallel hard cubes of side length three on a simple cubic lattice. A fully convolutional network learns the one-body direct-correlation functional $c^{(1)}[\rho]$ from data obtained with grand-canonical Monte Carlo simulations in randomized external potentials. Complete profiles are used during both training and inference; a stochastic Bernoulli mask on the output sites makes full-profile training effective without explicitly extracting and storing overlapping local density windows. Averaging the first-layer kernels over all 48 rotations and reflections of the cubic point group additionally imposes exact cubic equivariance without data augmentation. We compare the learned functional with independent simulation data and with the lattice fundamental-measure functional of Lafuente and Cuesta. The neural functional markedly improves the homogeneous equation of state and the density profile at a planar hard wall. For the anisotropic pair distribution around a fixed particle, both functionals reproduce the principal packing shells, with their relative accuracy depending on crystallographic direction. These results demonstrate neural density-functional calculations on complete three-dimensional profiles while also identifying accurate full-dimensional training data, thermodynamic consistency, and structural correlations as the central challenges for extensions to continuum fluids.
\end{abstract}

\author{Jens Weimar}
\email{jens.weimar@uni-tuebingen.de}
\affiliation{Institute of Applied Physics, University of T\"ubingen, Germany}

\author{Martin Oettel}
\affiliation{Institute of Applied Physics, University of T\"ubingen, Germany}

\author{Alessandro Simon}
\email{alessandro.simon@uni-tuebingen.de}
\affiliation{Institute of Applied Physics, University of T\"ubingen, Germany}

\maketitle

\section{Introduction}

Classical density functional theory (DFT) replaces the equilibrium many-body problem by a variational problem for the one-body density $\rho(\rr)$ \cite{bobstart,mermin1965thermal}. If the intrinsic excess free-energy functional were known exactly, the theory would provide equilibrium structure and thermodynamics in arbitrary external fields at a cost far below that of repeated particle simulations. Exact nontrivial functionals are, however, known only in special cases, most prominently the one-dimensional hard-rod fluid \cite{percusfunctional}. Fundamental measure theory (FMT) provides highly successful geometrically motivated approximations for continuum hard particles \cite{rosenfeldfunctional,rothsreview,whitebearmark2,lutskofunctional,melihsfunctional} and for hard-core lattice gases \cite{lafuente2002latticefmt,lafuentecuesta}, but its accuracy remains model and observable dependent.

Machine learning offers a complementary route in which the functional is inferred from equilibrium density profiles. Existing approaches range from direct representations of a free energy, mean-field kernel learning, to weighted-density, correlation-matching constructions and neural operator learning\cite{sammldft,sammldft2,catsroij,dftmlpaircorrmatching,kelley2024,reviewalessandromartin,ml-jctc,ml-paris,neuraloperator}. In the neural-functional method introduced by Samm\"uller \emph{et al.}, the network represents the one-body direct correlation $c^{(1)}$ rather than an explicit scalar free energy \cite{sammuellerc1,sammuellerc1howto}. The target follows directly from the Euler--Lagrange equation, so grand-canonical simulations in randomized external fields supply supervised training pairs without requiring free-energy labels. This strategy has achieved near-simulation accuracy for hard spheres \cite{sammuellerc1} and has since been applied to phase coexistence, mixtures, and ionic fluids \cite{sammuellerc1temperature,robitschkollps,buiionicfluidsc1}.

The original local-learning construction uses a multilayer perceptron (MLP) to map a fixed-size density window centered at one spatial point to the scalar $c^{(1)}$ at that point \cite{sammuellerc1,sammuellerc1howto}. Most subsequent implementations retained this MLP/window formulation \cite{sammuellerpaircorrresponse,sammuellerc1temperature,robitschkollps,kampa1,kampa2,kampa3,sammuellerhyperdensity,sammuellerhyperdensityhowto,sammuellermu,buiionicfluidsc1,buihyperdft1,buihyperdft2,buiunifiedmachinelearningframework,zhouc1azeotrope} because it is simple and flexible. For example, temperature and the pair potential can be supplied as additional inputs nodes \cite{sammuellerc1temperature,robitschkollps,kampa1,kampa2,kampa3}. However, it ties the network architecture to the data pipeline: a separate window must be extracted for every target point, and a complete $c^{(1)}$ profile is obtained by evaluating these windows and reassembling their scalar outputs.  This overhead is irrelevant for one-dimensional profiles but grows rapidly with dimension because both the number of windows and the number of density values in each window grow as powers of the spatial resolution.

Glitsch \emph{et al.} disentangled the local map from explicit window extraction by rewriting it as a fully convolutional network \cite{2dmldft}. One finite-range convolution evaluates the first MLP layer at every position, while subsequent $1\times1$ convolutions implement the remaining pointwise channel mixing. The resulting network can evaluate an entire profile efficiently. Crucially, however, a convolutional architecture does not require training on complete profiles. It was found that treating every site of a profile as one update amounts to an extremely large batch of strongly overlapping, correlated windows and gives an overly averaged, low-diversity gradient. Therefore sampled local windows during training were retained from the original formulation of Sammüller \emph{et al.} while the benefits of the convolutional form for full-profile inference could be used.

Here we retain both advantages in three dimensions: convolutional evaluation of complete fields and stochastic training updates, without ever constructing a window-expanded data set.  Every forward pass acts on several full $50 \times 50 \times 50$ profiles, but only an independently sampled Bernoulli subset of valid $c^{(1)}$ outputs contributes to the loss. The mask prevents each update from averaging over all spatial targets, while batching different profiles introduces variation across external fields and thermodynamic states.  This combination removes the memory and indexing costs of explicit three-dimensional windows and is the practical step that makes full-profile training feasible.  The lattice model retains excluded-volume correlations, packing oscillations, and direction-dependent pair structure while permitting sufficiently accurate training fields to be generated at manageable cost.

Our network uses one finite-range spatial convolution followed by a pointwise residual network. The first-layer kernels are projected onto the full cubic point group, making the density-to-$c^{(1)}$ map exactly equivariant under every lattice rotation and reflection. The invariant first-layer kernels compress each $7 \times 7 \times 7$ neighborhood into linear combinations of its 20 cubic-orbit density sums (see Appendix \ref{app:kernels}). This guarantees cubic equivariance but constitutes a stronger restriction than symmetry alone, because variations within each orbit are not retained.

We assess the learned functional against grand-canonical Monte Carlo (GCMC) data that are independent of the training profiles and against the analytical lattice FMT of Lafuente and Cuesta (LC). The comparisons cover the homogeneous equation of state, a planar hard-wall slit, and the full three-dimensional test-particle problem. The neural functional is substantially more accurate for the first two observables, whereas the pair-structure comparison is more nuanced.

\section{Theory and model}

\subsection{Lattice hard cubes}

We consider a simple cubic lattice $\Lambda$ of spacing $a$ and linear size $L$, with periodic boundary conditions unless stated otherwise. Particle orientations are fixed along the lattice axes. A particle centered at $\rr_0$ occupies the sites $\rr$ whose minimum-image coordinate differences satisfy
\begin{equation}
 \abs{r_\alpha-r_{0,\alpha}}_{\mathrm P}
 \leq \frac{s-1}{2},
 \qquad \alpha\in\{x,y,z\},
 \label{eq:cube-support}
\end{equation}
where the integer $s$ is the cube side in lattice units and $\abs{\cdot}_{\mathrm P}$ denotes a periodic minimum-image distance.  No lattice site may be occupied by more than one cube.  Consequently, two centers are incompatible if all three of their coordinate separations are at most $s-1$.  We use $s=3$ throughout, so a particle occupies $3^3=27$ sites and excludes the center of another particle from a $5 \times 5 \times 5$ region.  For a homogeneous center density $\rho$, we define the packing fraction
\begin{equation}
 \eta=s^3\rho.
 \label{eq:packing-fraction}
\end{equation}
Lengths are reported in units of $a$, and $k_{\mathrm B}T$ is the energy unit.

We note that the homogeneous fluid of $3\times3\times3$ lattice cubes has previously been studied using virial expansions and grand-canonical Monte Carlo simulations \cite{cubes1,cubes2,cubes3} and preliminary simulations have also addressed its high-density ordered phases \cite{cubes4}. We instead focus on the inhomogeneous, low to intermediate density state here.

The simulations sample the grand-canonical ensemble with particle insertion, deletion, and nearest-neighbor translation moves.  The external potential $V_{\mathrm{ext}}(\rr)$ acts on the particle center only. Overlaps are detected through an auxiliary site-occupancy field, while particle centers are stored separately for efficient translations. One-body densities are obtained by histogramming the center positions,
$\rho(\rr)=\avg{\hat\rho(\rr)}$.

\subsection{One-body direct-correlation functional}
For a lattice density field, the dimensionless grand-potential functional is
\begin{equation}
 \begin{split}
 \beta\Omega[\rho]={}&
 \sum_{\rr\in\Lambda}\rho(\rr)\bigl[\ln\rho(\rr)-1\bigr]
 +\beta F_{\mathrm{ex}}[\rho] \\
 &+\sum_{\rr\in\Lambda}\rho(\rr)
 \bigl[\beta V_{\mathrm{ext}}(\rr)-\beta\mu\bigr],
 \end{split}
 \label{eq:grand-potential}
\end{equation}
where the reference contribution to the chemical potential has been absorbed into $\mu$.
The one-body direct correlation functional is given by
\begin{equation}
    c^{(1)}(\rr;[\rho]) = -\beta \, \frac{\delta F_{\mathrm{ex}}[\rho]}{\delta\rho(\rr)}.
 \label{eq:c1-definition}
\end{equation}
Stationarity of Eq.~\eqref{eq:grand-potential} gives
\begin{equation}
 \rho(\rr)=\exp\!\left[
 \beta\mu-\beta V_{\mathrm{ext}}(\rr)
 +c^{(1)}(\rr;[\rho])\right].
 \label{eq:euler}
\end{equation}
For an equilibrium GCMC profile, Eq.~\eqref{eq:euler} can be inverted pointwise to obtain the supervised target
\begin{equation}
 c^{(1)}_{\mathrm{GCMC}}(\rr)
 =\ln\rho(\rr)+\beta V_{\mathrm{ext}}(\rr)-\beta\mu.
 \label{eq:training-target}
\end{equation}
The neural model approximates the functional map $\rho\mapsto c^{(1)}[\rho]$. Once trained, it is inserted into Eq.~\eqref{eq:euler} and solved self-consistently for a new external potential.

\subsection{Lafuente--Cuesta lattice functional}
\label{sec:lc}

The hard-core lattice model admits an FMT construction following Lafuente and Cuesta \cite{lafuente2002latticefmt,lafuentecuesta}.  We use this LC functional as the analytical reference in every comparison.  For monodisperse parallel cubes of odd side length $s=3$, its excess free energy is
\begin{equation}
 \beta F_{\mathrm{ex}}^{\mathrm{LC}}[\rho]
 =\sum_{\rr\in\Lambda}
 \sum_{\kk\in\{0,1\}^3}
 (-1)^{3-\abs{\kk}}
 \Phi_0\!\left(n^{(\kk)}(\rr)\right),
 \label{eq:lc-functional}
\end{equation}
where $\abs{\kk}=k_x+k_y+k_z$ and
\begin{equation}
 \Phi_0(\eta)=\eta+(1-\eta)\ln(1-\eta)
 \label{eq:phi-zero}
\end{equation}
is the exact excess free energy of a zero-dimensional cavity.  With the support convention used in our implementation, the eight weighted densities are
\begin{equation}
 n^{(\kk)}(x,y,z)
 =\sum_{i=0}^{1+k_x}
  \sum_{j=0}^{1+k_y}
  \sum_{\ell=0}^{1+k_z}
 \rho(x+i,y+j,z+\ell).
 \label{eq:lc-weighted-densities}
\end{equation}
All indices in Eq.~\eqref{eq:lc-weighted-densities} are periodic. The support extends in the positive coordinate directions only and therefore looks asymmetric about $(x,y,z)$.  This is only a choice of origin for the discrete weights: taking the functional derivative applies the adjoint weights and restores the full symmetry of $c^{(1)}_{\mathrm{LC}}$.  Other translated or reflected support conventions give the same functional.

The LC one-body direct correlation is calculated analytically from
$c^{(1)}_{\mathrm{LC}}=-\delta\beta F_{\mathrm{ex}}^{\mathrm{LC}}/\delta\rho$ and inserted into Eq.~\eqref{eq:euler}. For a uniform density, $n^{(\kk)}=v_{\kk}\,\rho$ with
$v_{\kk}=\prod_\alpha(2+k_\alpha)$, which also gives the LC bulk free energy and equation of state directly. Because Eq.~\eqref{eq:lc-functional} is an explicit scalar functional, its thermodynamic derivatives and mixed functional derivatives are internally consistent by construction.

\section{Simulation data and neural functional}

\subsection{Three-dimensional training fields}

The local-learning construction of Refs.~\cite{sammuellerc1,sammuellerc1howto} associates the value of $c^{(1)}$ at one site with a finite neighborhood of the density. For a field containing $L^3$ sites and a receptive field containing $K^3$ density values, pre-extracting all MLP inputs produces $L^3K^3$ stored values before targets and intermediate activations are included. In the present case, one $50 \times 50 \times 50$ profile would expand from 125,000 density values to 42,875,000 entries in its overlapping $7 \times 7 \times 7$ windows, a factor of 343. Generating only the requested windows on demand avoids this stored expansion but reintroduces repeated indexing and copying. Our fully convolutional network instead evaluates the same shared local map at every site directly from the unexpanded profile \cite{2dmldft}.

All training profiles have shape $50 \times 50 \times 50$.  Their smooth part is generated from periodic Gaussian fields
\begin{equation}
 \beta V_{\mathrm{smooth}}(\rr)
 =\sum_{i=1}^{N_G} A_i
 \exp\!\left[-\alpha_i
 \abs{\rr-\rr_i}_{\mathrm P}^2\right],
 \label{eq:random-potential}
\end{equation}
where the centers $\rr_i$ are uniform on the lattice and
$\alpha_i\in[10^{-3},1]$. The training corpus combines broad-amplitude Gaussian fields, narrower-amplitude fields containing random hard cuboids, and Gaussian fields inside a cavity closed by hard walls on all six faces. Table~\ref{tab:training-data} in the Appendix summarizes the four campaigns.  Combining these environments exposes the functional to bulk-like regions, smoothly varying fields, sharp exclusions, edges, and corners.

All campaigns use at least $4\times10^9$ production moves per replica and discard the first $10^8$ thermalization moves, and then sample the density every 100 moves.
In campaign C the hard cuboids have independently sampled side lengths from three to six sites. In campaign D five center-exclusion layers are placed at each face, leaving an accessible $40 \times 40 \times 40$ cavity. Fifty of the 443 complete profiles have mean packing fraction above $0.60$ and are excluded.  The final corpus therefore contains 393 profiles spanning mean packing fractions from approximately $0.063$ to $0.6$.

\subsection{Network architecture and cubic equivariance}
\label{sec:network}

The neural $c^{(1)}$ functional is a fully convolutional scalar-field network. Its first layer applies 32 kernels of size $7\times7\times7$, with unit stride and circular padding by three sites. The receptive field of every output value is consequently a $7 \times 7 \times 7$ density neighborhood.  A Leaky-ReLU activation with negative slope $0.05$ follows the spatial convolution. A $1\times1\times1$ convolution projects the 32 local features to 128 hidden channels. Three residual blocks then act pointwise; each block contains two $1\times1\times1$ convolutions with 128 input and output channels and Leaky-ReLU activations.  A final pointwise convolution produces one $c^{(1)}$ value per lattice site.  There is no pooling, normalization, or spatial downsampling.  The stored model contains 114,433 trainable parameters, though a large part of the kernel parameters are redundant because of the cubic symmetry.

This ordering is the fully convolutional counterpart of the original windowed MLP: the $7 \times 7 \times 7$ convolution applies the first dense layer to every local neighborhood in parallel, and the $1 \times 1 \times 1$ layers apply the remaining nonlinear map independently at every site.  

Glitsch \emph{et al.} also introduced an FMT-inspired alternative with a finite-range convolution both before and after the pointwise nonlinear network \cite{2dmldft}. The first convolution constructs learned weighted densities and the second performs an FMT-like back-convolution to $c^{(1)}$.
Although splitting a receptive field between two smaller kernels can reduce the forward kernel volume, the final finite-range convolution must backpropagate through a dense spatial convolution into every latent channel and then through the complete pointwise network, which increases the computational demand considerably with increasing dimensionality.
Since performance is similar for both architectures in the 2D case \cite{2dmldft}, we restrict ourself to the convolutional counterpart of the original formulation.

The hard-cube Hamiltonian is invariant under the full point group of the simple cubic lattice. We impose this symmetry on the neural map following the general principle of group-equivariant convolutional networks \cite{cohen2016group,kelley2024}. Let $G=\Oh$ be the 48-element cubic point group, represented on the lattice by all signed permutations of the three coordinate axes. Its action on a scalar density $T_g \rho$ is defined by
\begin{equation}
 (T_g\rho)(\rr)=\rho(g^{-1}\rr),
 \qquad g\in\Oh.
 \label{eq:group-action}
\end{equation}
Every raw first-layer kernel $K_q(\uu)$ is projected onto the $G$-invariant subspace,
\begin{equation}
 \overline K_q(\uu)
 =\frac{1}{\abs{G}}
 \sum_{g\in G}K_q(g^{-1}\uu).
 \label{eq:kernel-projection}
\end{equation}
In practice, Eq.~\eqref{eq:kernel-projection} is evaluated as the eight independent axis reflections followed by the six permutations of the axes during every forward pass. Convolution with $\overline K_q$ is equivariant under $G$. All subsequent operations act independently at each site through pointwise channel mixing, nonlinearities, and residual additions, and therefore commute with $T_g$. The complete model obeys
\begin{equation}
    c^{(1)}_\theta(\rr;[T_g\rho]) =c^{(1)}_\theta(g^{-1}\rr;[\rho]) =(T_gc^{(1)}_\theta[\rho])(\rr) \label{eq:equivariance}
\end{equation}
up to floating-point roundoff. Shared convolutional weights and circular padding additionally provide lattice-translation equivariance.

The distinction between invariance and equivariance is useful here.  Each effective spatial kernel, and hence the scalar response to a transformed local environment at its center, is invariant under $\Oh$.  Evaluating this response at every lattice site makes the complete density-field-to-$c^{(1)}$-field map equivariant.  This is a deliberately restricted scalar-channel construction, not a general steerable network\cite{weiler2018steerable} with orientation-indexed features, and it enforces only the discrete symmetry of the cubic lattice rather than arbitrary continuous rotations, which are naturally restricted by the lattice geometry.

\subsection{Training and model selection}

The 393 profiles are divided by a fixed random permutation into 354 training and 39 validation profiles. Each optimization batch contains ten complete fields, and Eq.~\eqref{eq:training-target} is evaluated on the fly. A single $50\times50\times50$ field supplies as many as 125,000 target sites, so an unmasked batch would average up to $1.25\times10^6$ local contributions.  These are not independent samples: neighboring outputs have strongly overlapping receptive fields, and all sites within one profile share the same chemical potential and external environment.  Averaging every site therefore produces a very large effective window batch with little stochastic variation and can wash out informative gradient differences between local environments.

We retain full-field convolution but stochastically subsample its outputs when constructing the loss.  Sites with $c^{(1)}_{\mathrm{GCMC}}\leq-10$ or nonfinite targets are first excluded, which removes hard-wall sites and regions with insufficient density statistics. At every optimization step, an independent Bernoulli mask with retention probability $p=0.5$ is then drawn for every remaining site. We can choose this comparatively large value because the corresponding density windows remain relatively small for the lattice system.
For continuous systems, $p$ needs to be chosen much smaller to prevent gradient flattening. If $\mathcal M$ denotes the selected profile--site pairs, the loss is
\begin{equation}
 \mathcal L(\theta)=\frac{1}{\abs{\mathcal M}}
 \sum_{(p,\rr)\in\mathcal M}
 \left[c^{(1)}_\theta(\rr;[\rho_p])
 -c^{(1)}_{p,\mathrm{GCMC}}(\rr)\right]^2.
 \label{eq:loss}
\end{equation}
The mask does not reduce the dense convolutional forward pass. Instead, its purpose is to provide a stochastic spatial estimate of the loss while preserving the memory and throughput advantages of unwindowed data. Batching ten independently generated profiles supplies the complementary variation between external fields and state points. Thus, the convolutional architecture, full-profile data loading, and spatial loss sampling are separate design choices. Validation is deterministic and uses every finite site above the same target cutoff.

We train in single precision for 400 epochs with Adam, learning rate $10^{-3}$, and weight decay $10^{-5}$. The checkpoint is updated whenever the validation loss decreases and the minimum-validation checkpoint is restored after training.  
This selection avoids the intermittent late-epoch validation spikes. A preliminary otherwise comparable model with 16 rather than 32 first-layer kernels gave unsatisfactory validation and downstream profile accuracy; all results below therefore use 32 kernels. The learned kernels and this channel-count observation are discussed further in the Appendix.

For inhomogeneous predictions, Eq.~\eqref{eq:euler} is solved by Picard iteration with mixing,
\begin{equation}
 \rho^{(n+1)}=(1-\gamma)\rho^{(n)}
 +\gamma\exp\!\left[\beta\mu-\beta V_{\mathrm{ext}}
 +c^{(1)}[\rho^{(n)}]\right],
 \label{eq:picard}
\end{equation}
with inaccessible sites reset to zero after each step.  The mixing parameter $\gamma$ is reduced when required for stability.  All reported ML and LC profiles satisfy their stated fixed-point convergence criteria.

\section{Results}

\subsection{Homogeneous equation of state}

For a constant density and zero external field, the neural Euler equation reduces to
\begin{equation}
 \beta\mu_{\mathrm{ML}}(\rho)=\ln\rho-c^{(1)}_{\mathrm{ML}}(\rho).
 \label{eq:ml-bulk-mu}
\end{equation}
We evaluate this relation on a dense density grid and invert its monotonic fluid branch.  The LC bulk relation follows analytically from Eq.~\eqref{eq:lc-functional}. Independent GCMC data comprise 20 chemical potentials between $\beta\mu=-8$ and $1.5$, eight replicas per state, and a periodic $48\times48\times48$ box commensurate with the particle side length.  Each replica contains $10^7$ density samples after thermalization.

Pressure is reconstructed consistently from
$\partial(\beta P)/\partial(\beta\mu)=\rho$. The integration is anchored at $\beta\mu=-8$ by the second-virial expression (accurate at low densities)
$\beta P=\rho+B_2\rho^2$, where
$B_2=(2s-1)^3/2$ for the present lattice cubes. GCMC pressure intervals are propagated by resampling the independent replicas before integration.

\begin{figure}
 \centering
 \includegraphics[width=\linewidth]{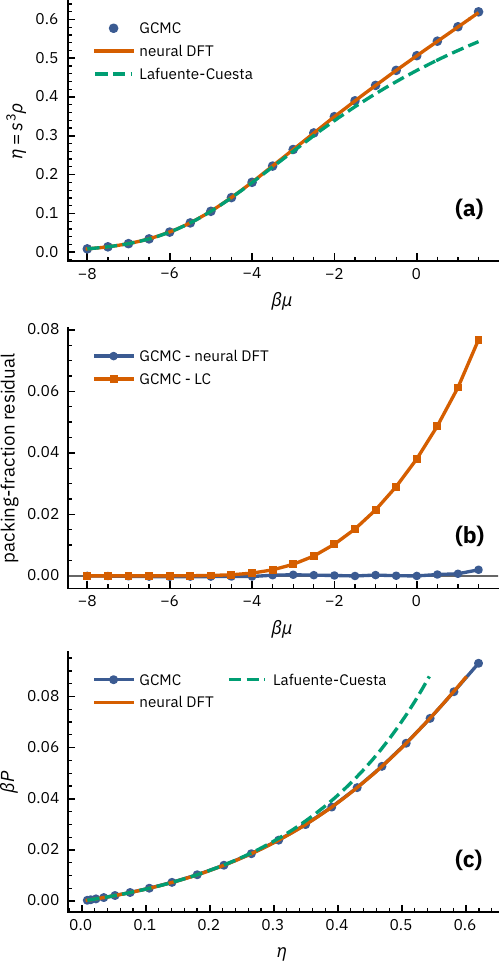}
 \caption{Homogeneous equation of state. (a) Packing fraction $\eta=s^3\rho$ as a function of chemical potential. (b) GCMC-minus-model packing-fraction residuals. (c) Pressure obtained by thermodynamic integration. The standard deviation for the GCMC data is smaller than the symbol size.}
 \label{fig:eos}
\end{figure}

Figure~\ref{fig:eos}(a) shows that the neural functional closely follows the simulated adsorption isotherm throughout the trained fluid range. The LC curve develops an increasing negative density bias with increasing chemical potential.  
The neural training-density boundary $\eta=0.60$ is reached at
$\beta\mu_{\mathrm{ML}}=1.265$. Values above it demonstrate the extrapolation capabilities of the learned network.
Figure~\ref{fig:eos}(c) demonstrates that the accuracy of the learned density relation carries over to the integrated pressure, whereas the LC pressure increasingly departs from the GCMC result at high packing fraction.

\subsection{Planar hard walls}

We next consider a pure planar slit that was not itself included among the training profiles. Infinite center potentials occupy five layers at each of the two faces normal to $z$, leaving 40 accessible layers in the periodic $50\times50\times50$ simulation box. The validation profile at $\beta\mu=0$ is averaged over 24 independent GCMC replicas, each with $2\times10^7$ density samples. Translational averaging over the two lateral directions and reflection averaging between the equivalent walls further reduce the simulation uncertainty.

\begin{figure}
 \centering
 \includegraphics[width=\linewidth]{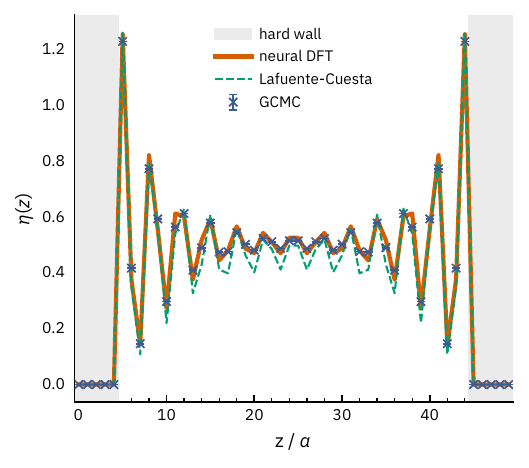}
 \caption{Packing-fraction profile in a planar hard-wall slit at $\beta\mu=0$.  The shaded regions are inaccessible to particle centers.  The neural and LC profiles are converged solutions of their Euler equations; symbols show the symmetry-averaged GCMC result from 24 replicas. The GCMC uncertainty is smaller than the symbol size.}
 \label{fig:hard-wall}
\end{figure}

All three profiles in Fig.~\ref{fig:hard-wall} exhibit the strong contact peak and damped packing oscillations induced by the walls.  The neural functional tracks the peak amplitudes and the oscillatory interior more accurately than LC.

\subsection{Pair distribution around a fixed particle}
\label{sec:pair-distribution}

The test-particle route provides a genuinely three-dimensional structural test \cite{percus1962approximation}.  Fixing one cube at the center creates an infinite external potential for mobile-particle centers in the surrounding $5\times5\times5$ exclusion region.  For a homogeneous bulk state of density $\rho_{\mathrm b}$, the resulting equilibrium profile gives the anisotropic pair distribution
\begin{equation}
 g(\rr)=\frac{\rho(\rr\mid\text{fixed particle})}{\rho_{\mathrm b}}.
 \label{eq:test-particle}
\end{equation}
Because the particles and lattice have only cubic symmetry, $g$ depends on the vector $\rr$ rather than on its magnitude alone.

The GCMC calculation uses a $50\times50\times50$ box, $\beta\mu=-2$, and eight paired bulk and fixed-particle replicas. It gives
$\eta_{\mathrm b}=0.345$.  To separate inhomogeneous structural errors from bulk equation-of-state errors, each functional is evaluated at the chemical potential that reproduces this same bulk density: $\beta\mu=-2.00$ for the neural model and $-1.86$ for LC. 

\begin{figure}
 \centering
 \includegraphics[width=0.95\linewidth]{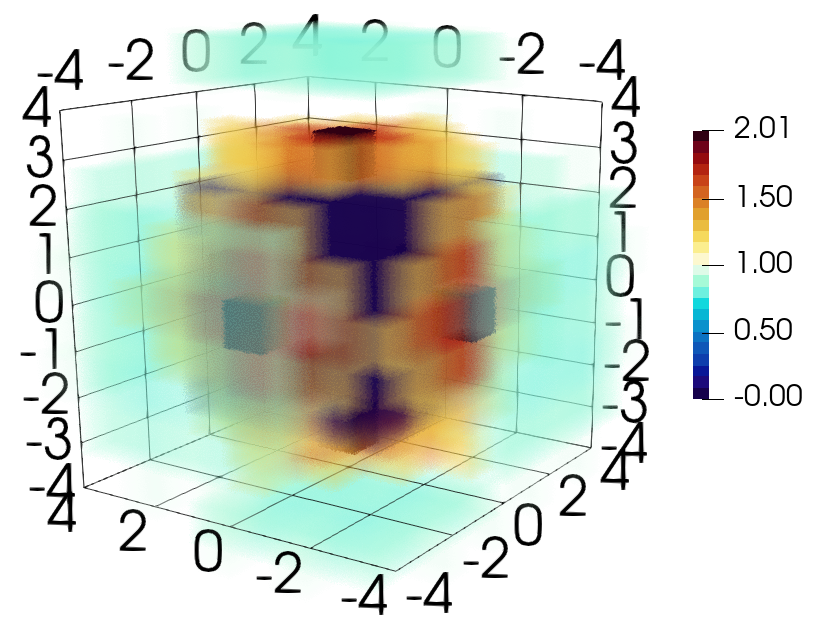}
 \caption{Volume rendering of the neural test-particle pair distribution $g(\rr)$ in the central region of the box.  The dark central cube is the $5\times5\times5$ region excluded to particle centers; the surrounding shells reveal the cubic anisotropy that is lost in a spherical average.  Quantitative directional cuts are shown in Fig.~\ref{fig:pair-cuts}.}
 \label{fig:pair-volume}
\end{figure}

\begin{figure*}
 \centering
 \includegraphics[width=\linewidth]{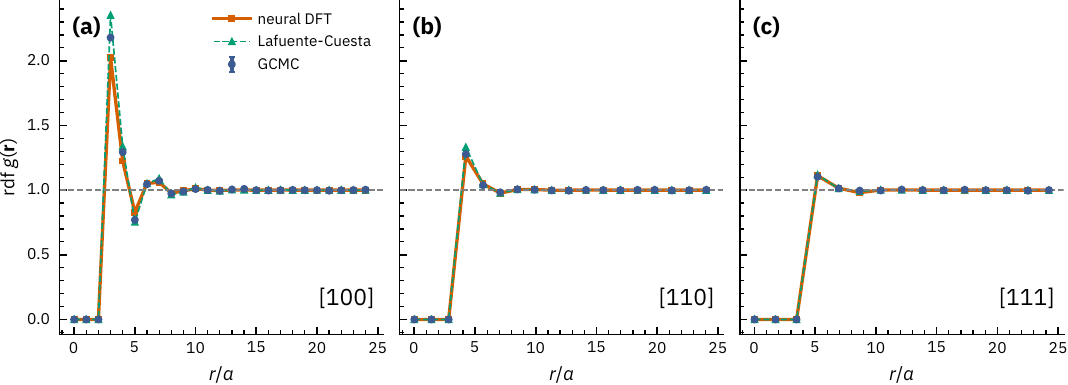}
 \caption{Directional pair distribution at matched bulk packing fraction $\eta_{\mathrm b}=0.3$: (a) $[100]$, (b) $[110]$, and (c) $[111]$.  GCMC symbols and standard errors are obtained from eight paired replicas after averaging all symmetry-equivalent rays.  Neural and LC calculations use chemical potentials adjusted separately to reproduce the GCMC bulk density.  The fixed particle excludes separations inside the central $5\times5\times5$ cube.}
 \label{fig:pair-cuts}
\end{figure*}

Figure~\ref{fig:pair-volume} illustrates the resulting three-dimensional neural profile, while Fig.~\ref{fig:pair-cuts} shows quantitative cuts along the three principal crystallographic families.  Signed and permutation-equivalent rays are averaged for every method.

The two functionals both capture the exclusion region, the first packing maximum, and the rapid decay toward $g=1$.  Their relative accuracy depends on direction.
Both methods describe the $[100]$ direction well, except for the contact value, which 
is slightly overestimated with LC and underestimated with the neural functional.
For the other directions the correlation
is rather small to begin with and well described in both cases, though the neural functional is marginally more accurate for $[110]$.
Unlike the bulk and wall results, this comparison does not support a claim of uniform neural superiority.  

\section{Conclusions and outlook}

We have constructed a neural approximation to the one-body direct-correlation functional of parallel hard cubes that acts on complete three-dimensional lattice fields.  The network combines a finite $7 \times 7 \times 7$ receptive field with pointwise residual channel mixing and an exact projection of the spatial kernels onto the 48-element cubic point group.  The local scalar response is invariant and the complete density-to-$c^{(1)}$ field map is equivariant under all lattice rotations and reflections.  This constraint prevents symmetry-equivalent environments from acquiring different learned responses.

The central computational step is the transition from an explicitly windowed MLP pipeline to full-profile convolution during both training and application.  Convolution alone resolves the cost of constructing and repeatedly evaluating overlapping windows, but it does not by itself provide useful stochastic training updates: an entire three-dimensional field represents a very large collection of correlated local environments.  Combining unexpanded profiles with Bernoulli subsampling of the output loss and batches containing several profiles separates efficient field evaluation from gradient sampling.  This preserves variation both within and between external environments and makes the approach practical at three-dimensional field sizes.

The learned functional reproduces the homogeneous GCMC equation of state with a packing-fraction RMSE of $1.99\times10^{-4}$ over the common supported range, compared with $1.36\times10^{-2}$ for the Lafuente--Cuesta functional.  
The fixed-particle test is more demanding: both functionals reproduce the principal anisotropic packing shells, but neither is uniformly more accurate across crystallographic directions.  These mixed structural results are important because they distinguish accurate self-consistent one-body profiles and bulk thermodynamics from a complete representation of pair correlations.

LC remains a valuable reference.  It is built from zero-dimensional consistency and dimensional crossover and, unlike the present network, is the derivative of an explicit excess free energy.  Directly learning $c^{(1)}$ does not guarantee that its functional Jacobian is symmetric or that a path-independent scalar $F_{\mathrm{ex}}$ exists.  The finite receptive field, the cutoff at mean packing fraction $0.60$, and the diversity and statistical precision of the training profiles provide additional controlled limitations.  Preliminary training with 16 first-layer kernels was inadequate, indicating that even after cubic symmetrization the present task requires a moderately rich local feature representation.

Although our demonstration uses a lattice fluid, the method is not restricted to lattice models. The same procedure can be applied to continuum fluids after representing their equilibrium densities on a spatial grid, provided the training data are sufficiently accurate and diverse.  In practice, obtaining smooth fully three-dimensional histograms, and therefore reliable logarithmic $c^{(1)}$ targets, is the principal bottleneck.  For an isotropic continuum fluid, the present $\Oh$ projection would impose only the cubic subgroup of the full rotational symmetry; an $O(3)$- or $E(3)$-equivariant steerable construction would be more appropriate but also more difficult to implement \cite{weiler2018steerable}.

Several extensions follow naturally.  Learning an equivariant scalar excess free energy would guarantee integrability, while pair-correlation matching or explicit sum-rule regularization could improve higher-order structure. Multiscale receptive fields could describe longer-ranged correlations without an excessively large dense kernel. Finally, extending the training set to ordered phases, mixtures, and more complex confinement would test whether the present three-dimensional strategy can retain its accuracy beyond the single-component fluid regime.

\section*{Acknowledgments}
This work is supported by the Deutsche Forschungsgemeinschaft (DFG, German Research Foundation), project 535083866.

\appendix

\section{Learned kernels and principal components}
\label{app:kernels}

\begin{figure}[h]
 \centering
 \includegraphics[width=\linewidth]{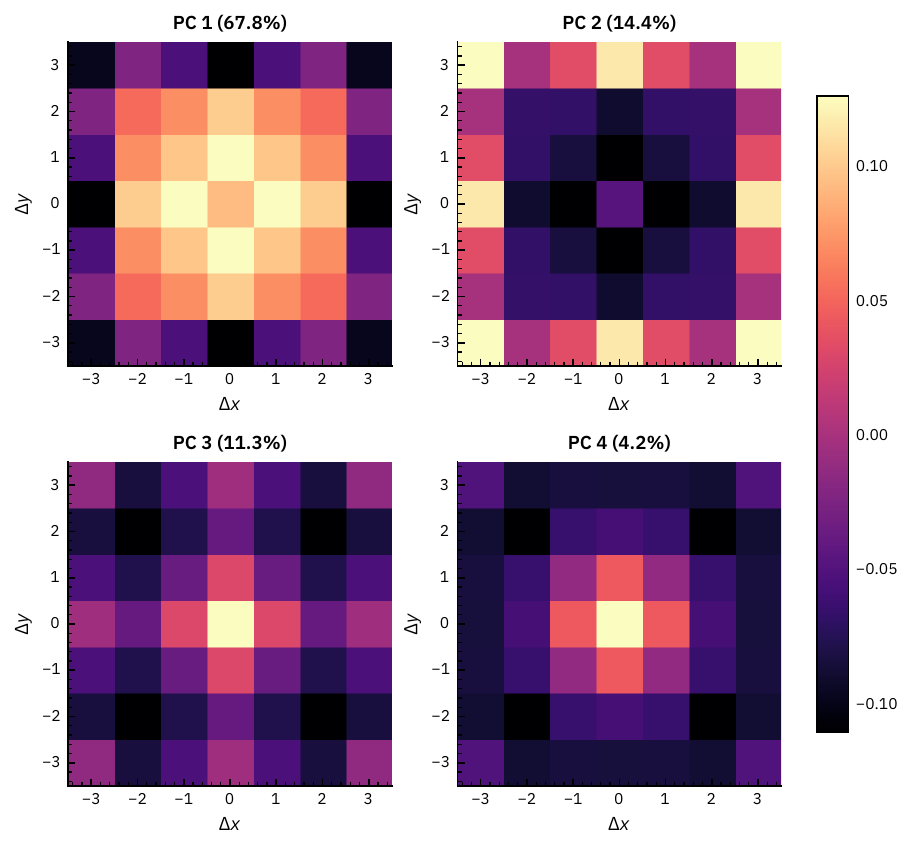}
 \caption{Central $xy$ cuts through the first four principal components of the 32 effective first-layer kernels.  Percentages give the fraction of centered kernel-to-kernel variance associated with each component.}
 \label{fig:kernel-pca}
\end{figure}

\begin{figure*}
 \centering
 \includegraphics[width=\linewidth]{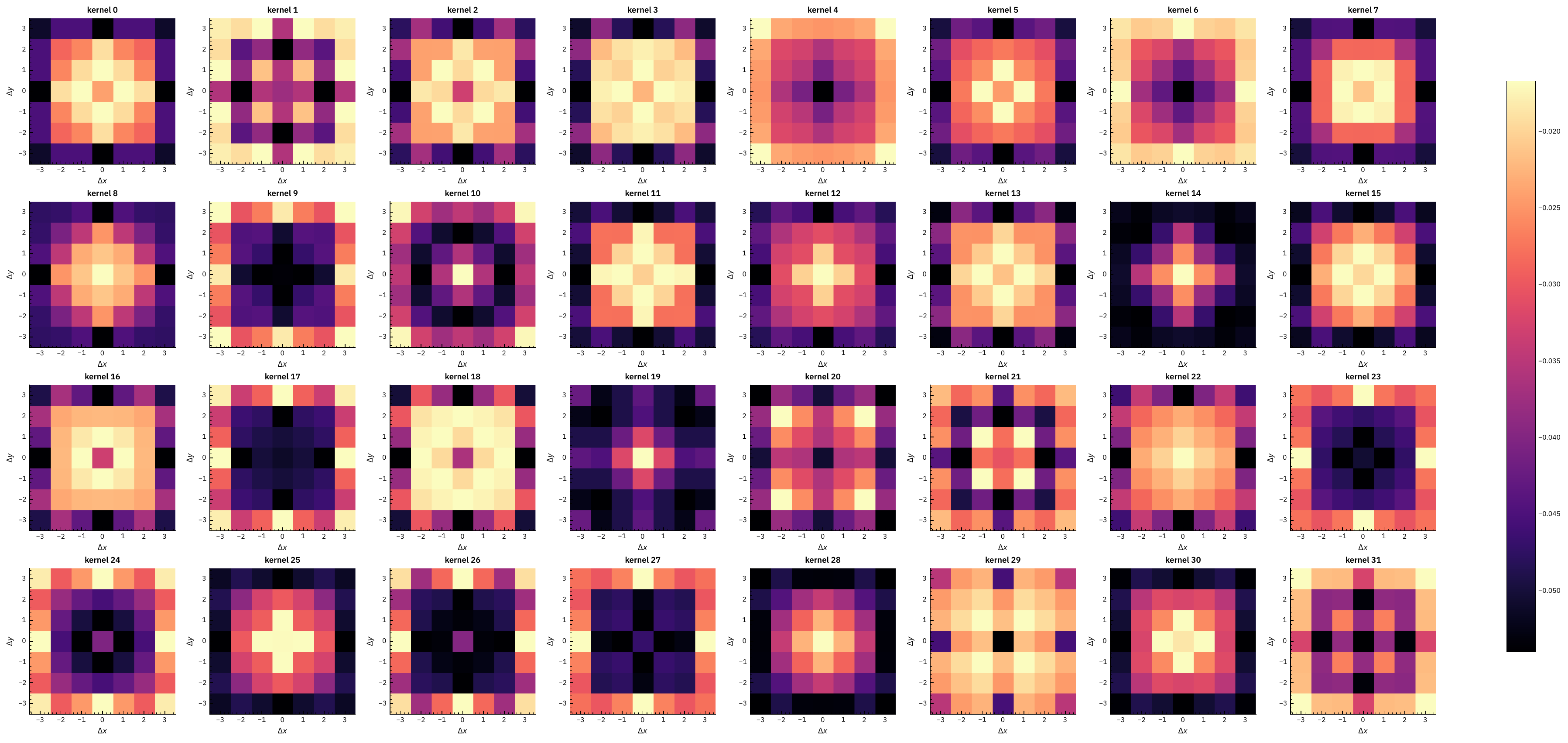}
 \caption{Central $xy$ cuts ($u_z=0$) through the 32 effective, cubic-symmetrized $7\times7\times7$ kernels.  The kernels are latent features optimized jointly with the pointwise network and should not individually be identified with FMT weight functions.  }
 \label{fig:kernels}
\end{figure*}

\begin{table*}[t]
 \caption{Composition of the training corpus.  Each density field is an average over 20 independent GCMC replicas.  Gaussian amplitudes are in units of $k_{\mathrm B}T$; all campaigns use $\alpha_i\in[10^{-3},1]$.}
 \label{tab:training-data}
 \begin{ruledtabular}
 \begin{tabular}{llllr}
 campaign & external environment & Gaussian parameters & sampled $\beta\mu$ \\
 \colrule
 A & periodic smooth field & $N_G=100$, $A_i\in[-3,10]$ & uniform in $[-2,3]$  \\
 B & periodic smooth field & $N_G=100$, $A_i\in[-3,10]$ & uniform in $[-2.5,1.8]$  \\
 C & smooth field and 10 hard cuboids & $N_G=50$, $A_i\in[-3,2]$ & uniform in $[-2.5,1.8]$  \\
 D & smooth field and six hard walls & $N_G=50$, $A_i\in[-3,2]$ & grid spanning $[-2.5,2.0]$ \\
 \end{tabular}
 \end{ruledtabular}
\end{table*}

The first convolution contains 32 effective kernels
$\overline K_q\in\mathbb R^{7\times7\times7}$.  Equation~\eqref{eq:kernel-projection} makes every one of them constant on orbits of the cubic point group.  For offsets with components in $\{-3,\ldots,3\}$, the 343 sites collapse to only 20 distinct orbits, labeled by sorted triples
$(\abs{u_x},\abs{u_y},\abs{u_z})$ with
$0\leq\abs{u_x}\leq\abs{u_y}\leq\abs{u_z}\leq3$.  The 32 filters are therefore different linear combinations of the same 20 symmetry-orbit sums.

To quantify common structure among the filters, we flatten each effective $7 \times 7 \times 7$ kernel to 343 entries, subtract the mean kernel across the 32 channels, and perform principal-component analysis by singular-value decomposition.  This voxel-space procedure retains the natural multiplicity of each cubic orbit.  The first four components account for $67.76\%$, $14.45\%$, $11.30\%$, and $4.17\%$ of the centered across-kernel variance, respectively, for a cumulative $97.68\%$.

The principal components in Fig.~\ref{fig:kernel-pca} are orthonormal directions in kernel space, not additional trained filters.  The cumulative $97.68\%$ means that most of the variation in first-layer kernel shapes lies in a four-dimensional linear subspace around their mean.  It does not mean that four components explain the same fraction of the network predictions, accuracy, or thermodynamic behavior; downstream weights and nonlinearities determine how strongly each filter affects $c^{(1)}$.  The PCA nevertheless suggests additional empirical compression beyond the exact reduction to 20 cubic orbits and motivates a controlled low-rank architecture study.

Figure~\ref{fig:kernels} shows the central $xy$ cut of each effective kernel.  The expected reflection and axis-exchange symmetries are visible in every panel, while the radial and shell-like variations differ across channels.  These kernels need not admit a unique physical interpretation: subsequent pointwise layers form nonlinear combinations of all 32 responses.  Nevertheless, the poor results of the preliminary 16-filter model are consistent with the need to combine a sufficiently varied set of local measurements.

\section{External potentials used for training data generation}

In Table~\ref{tab:training-data} we list the different kinds of external potentials that were used for generating the profiles used during training, including ranges of sampled values.

\bibliography{lit}

@article{sammuellerc1,
author = {Florian Sammüller  and Sophie Hermann  and Daniel de las Heras  and Matthias Schmidt },
title = {Neural functional theory for inhomogeneous fluids: Fundamentals and applications},
journal = {Proc. Natl. Acad. Sci.},
volume = {120},
number = {50},
pages = {e2312484120},
year = {2023},
doi = {10.1073/pnas.2312484120},
URL = {https://www.pnas.org/doi/abs/10.1073/pnas.2312484120}}

@article{sammuellerc1howto,
doi = {10.1088/1361-648X/ad326f},
url = {https://dx.doi.org/10.1088/1361-648X/ad326f},
year = {2024},
month = {mar},
publisher = {IOP Publishing},
volume = {36},
number = {24},
pages = {243002},
author = {Florian Sammüller and Sophie Hermann and Matthias Schmidt},
title = {Why neural functionals suit statistical mechanics},
journal = {J. Phys.: Condens. Matter}
}

@article{sammuellerc1temperature,
  title = {Neural Density Functional Theory of Liquid-Gas Phase Coexistence},
  author = {Samm\"uller, Florian and Schmidt, Matthias and Evans, Robert},
  journal = {Phys. Rev. X},
  volume = {15},
  issue = {1},
  pages = {011013},
  numpages = {23},
  year = {2025},
  month = {Jan},
  publisher = {American Physical Society},
  doi = {10.1103/PhysRevX.15.011013},
  url = {https://link.aps.org/doi/10.1103/PhysRevX.15.011013}
}

@article{sammuellerhyperdensity,
title = {Hyperdensity Functional Theory of Soft Matter},
author = {Samm\"uller, Florian and Robitschko, Silas and Hermann, Sophie and Schmidt, Matthias},
journal = {Phys. Rev. Lett.},
volume = {133},
issue = {9},
pages = {098201},
numpages = {7},
year = {2024},
month = {Aug},
publisher = {American Physical Society},
doi = {10.1103/PhysRevLett.133.098201},
url = {https://link.aps.org/doi/10.1103/PhysRevLett.133.098201}
}

@article{sammuellerhyperdensityhowto,
doi = {10.1088/1361-648X/ad98da},
url = {https://dx.doi.org/10.1088/1361-648X/ad98da},
year = {2024},
month = {dec},
publisher = {IOP Publishing},
volume = {37},
number = {8},
pages = {083001},
author = {Sammüller, Florian and Schmidt, Matthias},
title = {Why hyperdensity functionals describe any equilibrium observable},
journal = {J. Phys.: Condens. Matter}
}

@article{sammuellerpaircorrresponse,
title = {Neural density functionals: Local learning and pair-correlation matching},
author = {Samm\"uller, Florian and Schmidt, Matthias},
journal = {Phys. Rev. E},
volume = {110},
issue = {3},
pages = {L032601},
numpages = {6},
year = {2024},
month = {Sep},
publisher = {American Physical Society},
doi = {10.1103/PhysRevE.110.L032601},
url = {https://link.aps.org/doi/10.1103/PhysRevE.110.L032601}
}

@article{kampa1,
  title = {Metadensity Functional Theory for Classical Fluids: Extracting the Pair Potential},
  author = {Kampa, Stefanie M. and Samm\"uller, Florian and Schmidt, Matthias and Evans, Robert},
  journal = {Phys. Rev. Lett.},
  volume = {134},
  issue = {10},
  pages = {107301},
  numpages = {10},
  year = {2025},
  month = {Mar},
  publisher = {American Physical Society},
  doi = {10.1103/PhysRevLett.134.107301},
  url = {https://link.aps.org/doi/10.1103/PhysRevLett.134.107301}
}

@article{kampa2,
author = {Kampa, Stefanie M. and Samm{\"u}ller, Florian and Schmidt, Matthias},
title = {Metadensity Functional Learning for Classical Fluids: Regularizing with Pair Correlations},
journal = {The Journal of Physical Chemistry B},
volume = {130},
number = {24},
pages = {6231-6244},
year = {2026},
doi = {10.1021/acs.jpcb.6c01662},
note ={PMID: 42240207},
URL = {https://doi.org/10.1021/acs.jpcb.6c01662}
}

@article{kampa3,
  title={Spherical metadensity functional learning for inhomogeneous classical fluids},
  author={Kampa, Stefanie M and Schmidt, Matthias and Samm{\"u}ller, Florian},
  journal={arXiv preprint arXiv:2606.14370},
  year={2026}
}

@article{robitschkollps,
    author = {Robitschko, Silas and Sammüller, Florian and Schmidt, Matthias and Evans, Robert},
    title = {Learning the bulk and interfacial physics of liquid–liquid phase separation with neural density functionals},
    journal = {The Journal of Chemical Physics},
    volume = {163},
    number = {16},
    pages = {161101},
    year = {2025},
    month = {10},
    issn = {0021-9606},
    doi = {10.1063/5.0290261},
    url = {https://doi.org/10.1063/5.0290261},
}

@article{sammuellermu,
  title = {Determining the Chemical Potential via Universal Density Functional Learning},
  author = {Samm\"uller, Florian and Schmidt, Matthias},
  journal = {Phys. Rev. Lett.},
  volume = {136},
  issue = {6},
  pages = {068202},
  numpages = {9},
  year = {2026},
  month = {Feb},
  publisher = {American Physical Society},
  doi = {10.1103/7bqn-y2d7},
  url = {https://link.aps.org/doi/10.1103/7bqn-y2d7}
}

@article{buiionicfluidsc1,
  title = {Learning Classical Density Functionals for Ionic Fluids},
  author = {Bui, Anna T. and Cox, Stephen J.},
  journal = {Phys. Rev. Lett.},
  volume = {134},
  issue = {14},
  pages = {148001},
  numpages = {8},
  year = {2025},
  month = {Apr},
  publisher = {American Physical Society},
  doi = {10.1103/PhysRevLett.134.148001},
  url = {https://link.aps.org/doi/10.1103/PhysRevLett.134.148001}
}

@article{buihyperdft1,
doi = {10.1088/1361-648X/ade7e7},
url = {https://doi.org/10.1088/1361-648X/ade7e7},
year = {2025},
month = {jul},
publisher = {IOP Publishing},
volume = {37},
number = {28},
pages = {285101},
author = {Bui, Anna T and Cox, Stephen J},
title = {A first-principles approach to electromechanics in liquids},
journal = {Journal of Physics: Condensed Matter}
}

@article{buihyperdft2,
author={Bui, Anna T.
and Cox, Stephen J.},
title={Dielectrocapillarity for exquisite control of fluids},
journal={Nature Communications},
year={2026},
month={Feb},
day={12},
volume={17},
number={1},
pages={2661},
issn={2041-1723},
doi={10.1038/s41467-026-69482-1},
url={https://doi.org/10.1038/s41467-026-69482-1}
}

@article{buiunifiedmachinelearningframework,
author = {Anna T. Bui  and Stephen J. Cox },
title = {A unified machine-learning framework for ab initio multiscale modeling of liquids},
journal = {Proceedings of the National Academy of Sciences},
volume = {123},
number = {30},
pages = {e2610049123},
year = {2026},
doi = {10.1073/pnas.2610049123},
URL = {https://www.pnas.org/doi/abs/10.1073/pnas.2610049123}}

@article{zhouc1azeotrope,
author = {Zhou, Katie
L. Y. and Bui, Anna T. and Cox, Stephen J.},
title = {Roles of Bulk and
Surface Thermodynamics in the Selective
Adsorption of a Confined Azeotropic Mixture},
journal = {The Journal of Physical Chemistry B},
volume = {130},
number = {16},
pages = {4455-4466},
year = {2026},
month = {04},
issn = {1520-6106},
doi = {10.1021/acs.jpcb.6c00640},
url = {https://doi.org/10.1021/acs.jpcb.6c00640},
}

@article{2dmldft,
  title = {Neural density functional theory in higher dimensions with convolutional layers},
  author = {Glitsch, Felix and Weimar, Jens and Oettel, Martin},
  journal = {Phys. Rev. E},
  volume = {111},
  issue = {5},
  pages = {055305},
  numpages = {9},
  year = {2025},
  month = {May},
  publisher = {American Physical Society},
  doi = {10.1103/PhysRevE.111.055305},
  url = {https://link.aps.org/doi/10.1103/PhysRevE.111.055305}
}

@Inbook{reviewalessandromartin,
author="Simon, Alessandro
and Oettel, Martin",
editor="te Vrugt, Michael",
title="Machine Learning Approaches to Classical Density Functional Theory",
bookTitle="Artificial Intelligence and Intelligent Matter: Nanoscience, Soft Matter, Philosophy",
year="2026",
publisher="Springer Nature Switzerland",
address="Cham",
pages="83--113",
isbn="978-3-032-04129-6",
doi="10.1007/978-3-032-04129-6_6",
url="https://doi.org/10.1007/978-3-032-04129-6_6"
}

@article{ml-jctc,
    author = {Simon, Alessandro and Weimar, Jens and Martius, Georg and Oettel, Martin},
    title = {Machine Learning of a Density Functional for Anisotropic
Patchy Particles},
    journal = {Journal of Chemical Theory and Computation},
    volume = {20},
    number = {3},
    pages = {1062-1077},
    year = {2024},
    month = {01},
    issn = {1549-9618},
    doi = {10.1021/acs.jctc.3c01238},
    url = {https://doi.org/10.1021/acs.jctc.3c01238},
}

@article{ml-paris,
    author = {Simon, Alessandro and Belloni, Luc and Borgis, Daniel and Oettel, Martin},
    title = {The orientational structure of a model patchy particle fluid: Simulations, integral equations, density functional theory, and machine learning},
    journal = {The Journal of Chemical Physics},
    volume = {162},
    number = {3},
    pages = {034503},
    year = {2025},
    month = {01},
    issn = {0021-9606},
    doi = {10.1063/5.0248694},
    url = {https://doi.org/10.1063/5.0248694},
   
}

@article{sammldft,
title={{A classical density functional from machine learning and a convolutional neural network}},
author={Shang-Chun Lin and Martin Oettel},
journal={SciPost Phys.},
volume={6},
pages={025},
year={2019},
publisher={SciPost},
doi={10.21468/SciPostPhys.6.2.025},
url={https://scipost.org/10.21468/SciPostPhys.6.2.025},
}

@article{sammldft2,
author = {Lin, S.-C. and Martius, G. and Oettel, M.},
title = "{Analytical classical density functionals from an equation learning network}",
journal = {J. Chem. Phys.},
volume = {152},
number = {2},
pages = {021102},
year = {2020},
month = {01},
issn = {0021-9606},
doi = {10.1063/1.5135919},
url = {https://doi.org/10.1063/1.5135919},
}

@article{catsroij,
author = {Cats, Peter and Kuipers, Sander and de Wind, Sacha and van Damme, Robin and Coli, Gabriele M. and Dijkstra, Marjolein and van Roij, René},
title = {Machine-learning free-energy functionals using density profiles from simulations},
journal = {APL Mater.},
volume = {9},
number = {3},
pages = {031109},
year = {2021},
month = {03},
issn = {2166-532X},
doi = {10.1063/5.0042558},
url = {https://doi.org/10.1063/5.0042558},
}

@article{dftmlpaircorrmatching,
  title = {Learning Neural Free-Energy Functionals with Pair-Correlation Matching},
  author = {Dijkman, Jacobus and Dijkstra, Marjolein and van Roij, Ren\'e and Welling, Max and van de Meent, Jan-Willem and Ensing, Bernd},
  journal = {Phys. Rev. Lett.},
  volume = {134},
  issue = {5},
  pages = {056103},
  numpages = {7},
  year = {2025},
  month = {Feb},
  publisher = {American Physical Society},
  doi = {10.1103/PhysRevLett.134.056103},
  url = {https://link.aps.org/doi/10.1103/PhysRevLett.134.056103}
}

@article{kelley2024,
    author = {Kelley, Michelle M. and Quinton, Joshua and Fazel, Kamron and Karimitari, Nima and Sutton, Christopher and Sundararaman, Ravishankar},
    title = {Bridging electronic and classical density-functional theory using universal machine-learned functional approximations},
    journal = {The Journal of Chemical Physics},
    volume = {161},
    number = {14},
    pages = {144101},
    year = {2024},
    month = {10},
    issn = {0021-9606},
    doi = {10.1063/5.0223792},
    url = {https://doi.org/10.1063/5.0223792},
}

@article{neuraloperator,
author = {Pan, Runtong and Fang, Xinyi and Azizzadenesheli, Kamyar and Liu-Schiaffini, Miguel and Gu, Mengyang and Wu, Jianzhong},
title = {Neural operators for forward and inverse potential–density mappings in classical density functional theory},
journal = {The Journal of Chemical Physics},
volume = {163},
number = {16},
pages = {164120},
year = {2025},
month = {10},
issn = {0021-9606},
doi = {10.1063/5.0284515},
url = {https://doi.org/10.1063/5.0284515}
}

@article{mermin1965thermal,
  title = {Thermal Properties of the Inhomogeneous Electron Gas},
  author = {Mermin, N. David},
  journal = {Phys. Rev.},
  volume = {137},
  number = {5A},
  pages = {A1441--A1443},
  year = {1965},
  doi = {10.1103/PhysRev.137.A1441}
}

@article{bobstart,
author = {R. Evans},
title = {The nature of the liquid-vapour interface and other topics in the statistical mechanics of non-uniform, classical fluids},
journal = {Adv. Phys.},
volume = {28},
number = {2},
pages = {143--200},
year = {1979},
publisher = {Taylor \& Francis},
doi = {10.1080/00018737900101365},
URL = {https://doi.org/10.1080/00018737900101365}
}

@article{lafuentecuesta,
  title = {Density Functional Theory for General Hard-Core Lattice Gases},
  author = {Lafuente, Luis and Cuesta, Jos\'e A.},
  journal = {Phys. Rev. Lett.},
  volume = {93},
  issue = {13},
  pages = {130603},
  numpages = {4},
  year = {2004},
  month = {Sep},
  publisher = {American Physical Society},
  doi = {10.1103/PhysRevLett.93.130603},
  url = {https://link.aps.org/doi/10.1103/PhysRevLett.93.130603}
}

@article{lafuente2002latticefmt,
  title = {Fundamental Measure Theory for Lattice Fluids with Hard-Core Interactions},
  author = {Lafuente, Luis and Cuesta, Jos{\'e} A.},
  journal = {J. Phys.: Condens. Matter},
  volume = {14},
  number = {46},
  pages = {12079--12097},
  year = {2002},
  doi = {10.1088/0953-8984/14/46/314}
}

@article{percusfunctional,
author={Percus, J. K.},
title={Equilibrium state of a classical fluid of hard rods in an external field},
journal={Journal of Statistical Physics},
year={1976},
month={Dec},
day={01},
volume={15},
number={6},
pages={505-511},
issn={1572-9613},
doi={10.1007/BF01020803},
url={https://doi.org/10.1007/BF01020803}
}

@article{percus1962approximation,
  title = {Approximation Methods in Classical Statistical Mechanics},
  author = {Percus, J. K.},
  journal = {Phys. Rev. Lett.},
  volume = {8},
  pages = {462--463},
  year = {1962},
  doi = {10.1103/PhysRevLett.8.462}
}

@article{rothsreview,
doi = {10.1088/0953-8984/22/6/063102},
url = {https://dx.doi.org/10.1088/0953-8984/22/6/063102},
year = {2010},
month = {jan},
publisher = {},
volume = {22},
number = {6},
pages = {063102},
author = {Roland Roth},
title = {Fundamental measure theory for hard-sphere mixtures: a review},
journal = {J. Phys.: Condens. Matter}
}

@article{rosenfeldfunctional,
  title = {Free-energy model for the inhomogeneous hard-sphere fluid mixture and density-functional theory of freezing},
  author = {Rosenfeld, Yaakov},
  journal = {Phys. Rev. Lett.},
  volume = {63},
  issue = {9},
  pages = {980--983},
  numpages = {0},
  year = {1989},
  month = {Aug},
  publisher = {American Physical Society},
  doi = {10.1103/PhysRevLett.63.980},
  url = {https://link.aps.org/doi/10.1103/PhysRevLett.63.980}
}

@article{lutskofunctional,
  title = {Explicitly stable fundamental-measure-theory models for classical density functional theory},
  author = {Lutsko, James F.},
  journal = {Phys. Rev. E},
  volume = {102},
  issue = {6},
  pages = {062137},
  numpages = {14},
  year = {2020},
  month = {Dec},
  publisher = {American Physical Society},
  doi = {10.1103/PhysRevE.102.062137},
  url = {https://link.aps.org/doi/10.1103/PhysRevE.102.062137}
}

@article{whitebearmark2,
doi = {10.1088/0953-8984/18/37/002},
url = {https://doi.org/10.1088/0953-8984/18/37/002},
year = {2006},
month = {aug},
publisher = {},
volume = {18},
number = {37},
pages = {8413},
author = {Hansen-Goos, Hendrik and Roth, Roland},
title = {Density functional theory for hard-sphere mixtures: the White Bear version mark
II},
journal = {Journal of Physics: Condensed Matter}
}

@article{melihsfunctional,
  title = {Using test particle sum rules to improve approximations in classical density functional theory: White-Bear and White-Bear mark II versions of the Lutsko functional},
  author = {G\"ul, Melih and Roth, Roland and Evans, Robert},
  journal = {Phys. Rev. E},
  volume = {113},
  issue = {3},
  pages = {034104},
  numpages = {8},
  year = {2026},
  month = {Mar},
  publisher = {American Physical Society},
  doi = {10.1103/3vg8-mbf4},
  url = {https://link.aps.org/doi/10.1103/3vg8-mbf4}
}

@inproceedings{cohen2016group,
  title = {Group Equivariant Convolutional Networks},
  author = {Cohen, Taco and Welling, Max},
  booktitle = {Proceedings of the 33rd International Conference on Machine Learning},
  series = {Proceedings of Machine Learning Research},
  volume = {48},
  pages = {2990--2999},
  year = {2016},
  publisher = {PMLR},
  url = {https://proceedings.mlr.press/v48/cohenc16.html}
}

@inproceedings{weiler2018steerable,
  title = {{3D} Steerable {CNNs}: Learning Rotationally Equivariant Features in Volumetric Data},
  author = {Weiler, Maurice and Geiger, Mario and Welling, Max and Boomsma, Wouter and Cohen, Taco S.},
  booktitle = {Advances in Neural Information Processing Systems},
  volume = {31},
  year = {2018},
  publisher = {Curran Associates, Inc.},
  url = {https://proceedings.neurips.cc/paper/2018/hash/488e4104520c6aab692863cc1dba45af-Abstract.html}
}

@article{cubes1,
author  = {Mitra, S. K. and Allnatt, A. R.},
title   = {Approximate and Exact Virial Coefficients for Hard Rectangular Parallelepipeds and Cubes on a Simple Cubic Lattice},
journal = {The Journal of Chemical Physics},
volume  = {71},
number  = {5},
pages   = {2105--2108},
year    = {1979},
doi     = {10.1063/1.438582}
}

@article{cubes2,
author  = {Davis, Jonathan R. and Piccarreta, Michael V. and
           Rauch, Rory B. and Vanderlick, T. Kyle and
           Panagiotopoulos, Athanassios Z.},
title   = {Phase Behavior of Rigid Objects on a Cubic Lattice},
journal = {Industrial \& Engineering Chemistry Research},
volume  = {45},
number  = {16},
pages   = {5421--5425},
year    = {2006},
doi     = {10.1021/ie051041c}
}

@article{cubes3,
    author  = {Fernandes, Heitor C. Marques and Levin, Yan and
               Arenzon, Jeferson J.},
    title   = {Equation of State for Hard-Square Lattice Gases},
    journal = {Physical Review E},
    volume  = {75},
    pages   = {052101},
    year    = {2007},
    doi     = {10.1103/PhysRevE.75.052101}
}

@article{cubes4,
    author  = {Vigneshwar, N. and Mandal, Dipanjan and Damle, Kedar
               and Dhar, Deepak and Rajesh, R.},
    title   = {Phase Diagram of a System of Hard Cubes on the
               Cubic Lattice},
    journal = {Physical Review E},
    volume  = {99},
    pages   = {052129},
    year    = {2019},
    doi     = {10.1103/PhysRevE.99.052129}
}

\end{document}